\documentclass[]{aa}
\usepackage{graphicx,longtable,lscape}
\usepackage{txfonts,psfig}
\usepackage[]{natbib}
\def\ltsima{$\; \buildrel < \over \sim \;$}
\def\lsim{\lower.5ex\hbox{\ltsima}}
\def\gtsima{$\; \buildrel > \over \sim \;$}
\def\gsim{\lower.5ex\hbox{\gtsima}}

\newcommand{\be}{\begin{equation}}
\newcommand{\en}{\end{equation}}

\begin{document}
  \title{Timing accuracy of the Swift X-Ray Telescope in WT mode}
   \author{G.\ Cusumano\inst{1}, V.\ La Parola\inst{1}, M.\ Capalbi\inst{2,3}, M.\
   Perri\inst{2,3}, A. P.\ Beardmore\inst{4}, D. N.\ Burrows\inst{5}, S.\ Campana\inst{6}, 
   J. A.\ Kennea\inst{5}, J. P.\ Osborne\inst{4}, B. Sbarufatti\inst{5,6}, G.\ Tagliaferri\inst{6}}

   \offprints{G. Cusumano, cusumano@ifc.inaf.it}
   \institute{INAF -- Istituto di Astrofisica Spaziale e Fisica Cosmica di Palermo, 
        Via U.\ La Malfa 153, 90146 Palermo, Italy 
 \and
   ASI Science Data Center, Via G. Galilei, 00044 Frascati, Italy 
 \and
   INAF -- Osservatorio Astronomico di Roma, via Frascati 33, 00040 Monteporzio Catone, Italy
 \and
  Department of Physics and Astronomy, University of Leicester, University Road, Leicester LE1 7RH, UK
  \and
   Department of Astronomy and Astrophysics, The Pennsylvania State University, 525 Davey Laboratory, University Park, Pennsylvania 16802, USA
  \and
          INAF -- Osservatorio Astronomico di Brera, Via Bianchi 46, 23807 Merate, Italy }

\abstract
{The X-Ray Telescope (XRT) on board Swift was mainly designed to provide detailed position, timing and spectroscopic 
information on Gamma-Ray Burst (GRB) afterglows. During the mission lifetime  the fraction of observing time allocated
to other types of source has been steadily increased.
} 
{In this paper, we report on the results of the in-flight calibration of the
timing capabilities of the XRT in Windowed Timing read-out mode.}
{We use observations of the Crab pulsar to evaluate the accuracy of the pulse period 
determination by comparing the values obtained by the XRT timing analysis with the 
values derived from radio monitoring.

We also check the absolute time
reconstruction measuring the phase position of the main peak in the Crab profile and comparing it both with the value
reported in literature and with the result that we obtain from a simultaneous Rossi X-Ray Timing Explorer (RXTE)
observation.}
{We find that the accuracy in period determination for the Crab pulsar is of the order of a few picoseconds for the 
observation
with the largest data time span. The absolute time reconstruction, measured using the position of the Crab main 
peak, shows that the main peak anticipates the phase of the position reported in
literature for RXTE by $\sim 270$ $\mu$s on average ($\sim 150$ $\mu$s when data are reduced with the attitude file 
corrected with the  UVOT data). 
The analysis of the simultaneous Swift-XRT and RXTE Proportional Counter Array
(PCA) observations confirms that the XRT
Crab profile leads the PCA  profile by $\sim 200$ $\mu$s.
The analysis of XRT Photodiode mode data and BAT event data shows a main peak position in good agreement 
with the RXTE, suggesting 
the discrepancy observed in XRT data in Windowed Timing mode
is likely due to a systematic offset in the time assignment for this XRT read out mode.}
{}
\authorrunning {G.\ Cusumano et al.}
\titlerunning {      }

\maketitle

\begin{table*}
\scriptsize
\begin{center}
\begin{tabular}{l r r r c c c c r  } \hline
 Obs ID 	  & $T_{elapsed}$   &Exposure& snapshot &$T_{epoch}$ (TDB) &$P~^a$ &$\Delta P~^b$  &$P_{Radio}$&$P-P_{radio}$ \\   
		  &  (s)	    & (s)    &  $\#$		&  (MJD)	   & (ms)      & (ms)	   &	(ms)   & (ns)	      \\   \hline	     
  00050100007	  &6341.9		&1079.6  &2  &53395.0964704775688 &33.5724117 & 7.9$\times10^{-6}$ &33.5724126208 &-0.8   \\ 
 00050100016	  &24218.2		&7872.3  &5  &53454.2110966913096 &33.5745613 & 2.2$\times10^{-6}$ &33.5745609884 &0.3  \\ 
 00058970001	  &272341.0		&14049.7 &13 &53685.5802279020791 &33.5829830 & 1.4$\times10^{-6}$ &33.5829688914 &14.1  \\ 
 00050100019	  &17910.7		&2277.4  &4  &53804.1732743901957 &33.5872787 & 3.8$\times10^{-6}$ &33.5872782420 &0.5      \\ 
 00050100020	  &87377.5		&6602.1  &12 &53842.4607787839923 &33.5886667 & 2.4$\times10^{-6}$ &33.5886694548 &-2.7      \\ 
 00050100021	  &23920.8		&4068.7  &5  &54179.6713066744996 &33.6009160 & 2.6$\times10^{-6}$ &33.6009213337 &-5.3  \\ 
 00058990007	  &417.1		&417.2   &1  &54180.8642858685562 &33.6009330 & 1.5$\times10^{-5}$ &33.6009646766 &-32.0  \\ 
 00050100025	  &47091.4		&4408.4  &6  &54389.3152526219274 &33.6085555 & 4.4$\times10^{-6}$ &33.6085374589 &18.0   \\ 
 00050100026	  &86544.1		&2188.6  &10 &54407.4404817774775 &33.6091754 & 1.4$\times10^{-5}$ &33.6091959069 &-20.4    \\ 
 00050100027	  &167615.6		&2939.5  &11 &54408.9766301068885 &33.6092231 & 1.4$\times10^{-5}$ &33.6092517043 &-28.5   \\ 
 00050100028	  &183.0		&183.0   &1  &54410.5885580012327 &33.6092286 & 4.3$\times10^{-5}$ &33.6093102548 &-81.6    \\ 
 00050100029	  &51788.6		&664.6   &10 &54698.3010557031885 &33.6197938 & 5.2$\times10^{-5}$ &33.6197609238 &32.9  \\ 
 00050100030	  &52479.5		&9384.8  &10 &54698.3067390939323 &33.6197619 & 1.4$\times10^{-6}$ &33.6197611302 &0.7      \\ 
 00058990008	  &34342.2		&1961.8  &3 &54855.1593480434822 &33.6254844 & 7.7$\times10^{-6}$ &33.6254579123 &26.5  \\ 
 00058990009$\dag$	  &18154.6		&2661.1  &4 &54860.8810891913527 &33.6256511 & 9.6$\times10^{-6}$ &33.6256657087 &-14.6     \\ 
 00058990010	 	  &35558.6		&6188.7  &7 &55091.7671635479928 &33.6340536 & 1.3$\times10^{-6}$ &33.6340503694 &3.2	     \\ 
 00058990011$\dag$	  &29024.1		&1720.6  &6 &55221.3119214096732 &33.6387293 & 1.1$\times10^{-5}$ &33.6387543790 &-25.1    \\ 
 00058990013$\dag$	  &1043.0		&1043.0  &1 &55226.2780278549253 &33.6389076 & 7.4$\times10^{-6}$ &33.6389346895 &-26.9    \\ 
 00030371002$\dag$	  &982.0		&982.0   &1 &55461.7032943207232 &33.6474857 & 1.9$\times10^{-6}$ &33.6474823886 &3.2	     \\ 
 00030371003$\dag$	  &4182.2		&952.2   &2 &55462.8191109598702 &33.6475174 & 5.5$\times10^{-6}$ &33.6475228989 &-5.5       \\ 
 00030371004$\dag$	  &1012.0		&682.6   &2 &55463.5876044134566 &33.6475470 & 3.7$\times10^{-6}$ &33.6475507994 &-3.8       \\ 
 00030371005$\dag$	  &975.8		&975.8   &1 &55463.0517034763325 &33.6475465 & 1.5$\times10^{-5}$ &33.6475313433 &15.1       \\ 
 00030371006$\dag$	  &4170.3		&958.8   &2 &55463.2863167333053 &33.6475527 & 6.9$\times10^{-6}$ &33.6475398610 &12.9       \\ 
 00030371007$\dag$	  &978.6		&978.6   &1 &55463.8446768862923 &33.6475631 & 5.6$\times10^{-6}$ &33.6475601325 &2.9	    \\ 
 00030371008$\dag$	  &6083.2		&963.3   &2 &55464.1003792808988 &33.6475533 & 5.3$\times10^{-6}$ &33.6475694159 &-16.1    \\ 
 00030371012$\dag$	  &48177.7		&4345.0  &3 &55465.3235006383620 &33.6476146 & 5.1$\times10^{-6}$ &33.6476138216 &0.8	    \\ 
 00030371013$\dag$	  &70241.8		&4996.4  &4 &55466.4693566183778 &33.6476498 & 2.4$\times10^{-6}$ &33.6476554220 &-5.7      \\ 
 00030371014$\dag$	  &23738.6		&1902.4  &2 &55467.1899126402641 &33.6476776 & 4.4$\times10^{-6}$ &33.6476815817 &-4.0       \\ 
 00030371015$\dag$	  &1035.7		&1035.7  &1 &55471.0121225304683 &33.6478160 & 3.7$\times10^{-6}$ &33.6478203235 &-4.3      \\ 
 00058990012$\dag$	  &955.1		&955.1   &1 &55574.5999367492113 &33.6515869 & 1.8$\times10^{-6}$ &33.6515809950 &5.9	      \\
 00030371016$\dag$	  &1818.7		&1818.7  &1 &55663.7859681778500 &33.6548202 & 1.8$\times10^{-5}$ &33.6548186029 &1.6	     \\ 
 00030371017$\dag$	  &983.7		&983.7   &1 &55666.8520923440810 &33.6549298 & 5.6$\times10^{-6}$ &33.6549299033 &-0.1       \\ 
 00030371018$\dag$	  &989.8		&989.8   &1 &55667.7889097694278 &33.6549750 & 1.1$\times10^{-5}$ &33.6549639096 &11.1       \\ 
 00030371019$\dag$	  &981.8		&981.8   &1 &55668.6640721579242 &33.6549899 & 5.6$\times10^{-6}$ &33.6549956777 &-5.7  	 \\ 
 00030371020$\dag$	  &975.6		&975.6   &1 &55669.5940489891800 &33.6550312 & 3.7$\times10^{-6}$ &33.6550294354 &1.8		 \\ 
 00030371021$\dag$	  &1191.7		&1191.7  &1 &55670.0763377977928 &33.6550446 & 5.6$\times10^{-6}$ &33.6550469423 &-2.4  	 \\ 
 00030371023$\dag$	  &1440.6		&1427.6  &2 &55671.2719624312667 &33.6550879 & 5.6$\times10^{-6}$ &33.6550903427 &-2.5  	 \\ 
 00030371024$\dag$	  &1289.2		&1289.2  &1 &55672.8748742345924 &33.6551479 & 5.6$\times10^{-6}$ &33.6551485270 &-0.7  	 \\ 
 00050100031$\star$	  &1017.8  	        &1017.8  &1 &55776.9641634614309 &33.6589272 & 3.7$\times10^{-6}$ &33.6589269212 &0.3	    \\ 
 00058990014$\dag$	  &319.7		&319.7   &1 &55946.7126451355507 &33.6650358 & 2.6$\times10^{-5}$ &33.6650883028 &-52.4   \\
 00058990015$\dag$	  &1473.0		&1473.0  &1 &55951.3198558431759 &33.6652550 & 5.6$\times10^{-6}$ &33.6652555472 &-0.5      \\ \hline
\end{tabular}
\caption{Observation log and period search results. 
 }
\begin{list}{}{}
\item [$^a$]  Best period obtained through the folding technique.
\item [$^b$]  68\% confidence level error on the best period.  
\item [$\dag$] Observations where an attitude file corrected by UVOT data is
available. 
\item [$\star$] Observation simultaneous with the RXTE-PCA observation.
\end{list}
\label{log}
\end{center}
\end{table*}

\section{Introduction\label{intro} }

The Swift satellite \citep{gehrels04}, launched on 2004 Nov 20, was mainly designed to detect and
localize gamma-ray bursts (GRBs), providing autonomous rapid-response observations and
long-term monitoring of the afterglow emission in a broad energy band from UV/optical to X-rays.
After 7 years of operation it has observed more than 600 GRBs, collecting a huge amount
of prompt and afterglow data whose study has revolutionized our understanding of the GRB phenomenology.
Besides GRBs, Swift has observed many other classes of cosmic source, dedicating a 
growing fraction of its observing time to the so called ``secondary'' science. 

The Swift payload includes three instruments: the Burst Alert Telescope (BAT: \citealp{barthelmy05}) 
observing in the 15--350 keV band, the X-Ray Telescope (XRT; \citealp{burrows05})  with an
energy range of 0.3--10 keV and the Ultra-Violet/Optical Telescope (UVOT; \citealp{roming05})
 with a wavelength range of 170--650 nm.  
XRT is a focusing X-ray telescope based on  grazing incidence Wolter I mirrors, 
equipped with a single e2v CCD-22 detector at its focal plane. 
The XRT focal detector was designed to support four different data read-out modes (see \citealp{hill04} for an
exhaustive description) in order to cover a wide
range of source intensities and to follow rapid variability of transient phenomena through an automatic 
transition among the read-out modes according to the intensity of the source.
The Imaging read-out mode is used soon after the slew to a newly detected transient
phenomenon to record a 
prompt image of the field of view for a rapid evaluation of the source coordinates, with no spectroscopy or
 time resolved data.
The Photodiode mode (PD) was designed to observe very bright sources, up to 60 Crab, with no spatial information and to achieve 
a high timing resolution of 0.14 ms. This mode  has been disabled following a
micrometeorite hit on May 27, 2005 that produced a very high background rate due to hot pixels which cannot be avoided
during read-out in this mode \citep{abbey06}.
The Photon Counting  mode preserves the full imaging and spectroscopic resolution but with limited time
resolution (2.5 s).  
This operation mode is suitable for source fluxes below $\sim$1 mCrab.

Windowed Timing (WT) mode uses the central 200 columns of the CCD (corresponding to a 8 arcminutes wide detector strip) 
and provides a one dimensional image by compressing them into a single row. 
The timing resolution achieved by the WT read-out mode is 1.7791 ms. 
This operation mode is suitable for source fluxes below $\sim$1 mCrab.
This mode is now used for bright sources and, among the presently operational XRT read-out modes, is the one 
with the best timing resolution. 

In this paper we describe the in-flight timing calibration of the XRT WT mode, investigating 
the accuracy in the period search and the absolute timing through the analysis of the Crab 
pulsar data.
The Crab pulsar (PSR~J0534+2200) has been used for in-flight calibration for several X-ray telescopes (e.g.
\citealp{kuster02,rots04,kirsch04,terada08,molkov10}) and it is continuously monitored in the radio
with its ephemeris regularly updated on a monthly basis and made available by
the Jodrell Bank Observatory\footnote{http://www.jb.man.ac.uk/pulsar/crab.html}. The
shape of its pulse profile, with a prominent main peak whose position in X-rays (as measured by RXTE in
the energy band 2--16 keV) leads
the radio one by 344$\pm$40 $\mu$s ($0.0102\pm0.0012$ in phase, \citealp{rots04}), is a useful
tool to investigate the absolute timing performance of the XRT instrument.

The paper is organized as follows: Section 2 describes how time is assigned to photon events
in WT mode; Section 3 summarizes the Crab pulsar observations, the reduction of the XRT data
and of the Crab observation performed by RXTE simultaneously with XRT; Section 4 describes
the timing analysis and reports the results; in Section 5 we give a brief discussion of the
results.

\section{Reconstruction of photon arrival times in WT mode}

The XRT WT mode is a CCD read-out mode used to achieve high resolution
timing (1.7791 ms) with 1-D position information along the detector X direction.
In this mode only the central part of the CCD is used. Different ``windows'' can be set and 
for flight observations a 200 columns wide window (covering about 8 arcminutes)
is being operated. 
The read-out data are generated by clocking 10 rows (parallel
transfers) into the serial register, where the signal is summed, and then reading out the
central 200 columns.

Because in this mode the CCD rows are clocked at a regular rate (instead of transferring
charge into the store area and then reading it out, typical of the 2D frame transfer in PC mode), 
there is no inherent frame structure to the data stream.  For convenience, telemetry frames containing 
600 serial read-out rows (each one containing ten CCD rows) are 
generated when formatting the data. For each frame the header includes, for the first and last row, 
the time at which the last pixel in the row was digitized. In the following these two time values 
are called Frame Start Time (FST) and Frame End Time (FET), respectively. The values of FST and FET are 
expressed in MET (Mission Elapsed Time since January 1, 2001, UTC) and stored in the columns 'FSTS', 'FSTSub', 'FETS' and 'FETSub' 
of the XRT Housekeeping Header Packets file with a time resolution of 20 $\mu$s.
 
The temporal delays between the photon arrival times and the read-out times
are calculated during the ground processing.
This is done by a specific software 
module, named {\sc xrttimetag} which is part of the XRT Data Analysis Software 
({\sc XRTDAS}\footnote{http://heasarc.gsfc.nasa.gov/docs/swift/analysis/xrt\_swguide\_v1\_2.pdf})

The 'xrttimetag' task calculates the arrival time of each photon based on the number of row 
transfers and column shifts required to move that pixel from the row centered on the X-ray source  
(DETY) to the output register. 
More specifically, the photon arrival times (in MET) are reconstructed according to the following equation:

\begin{equation}
  t_{ph} = FST + RAWY \Delta t_{row} - \Delta t - 0.5 \Delta t_{row} 
\end{equation}

\noindent

where RAWY is the frame row number (0,599). The $\Delta t$ term is the transfer time to reach the output serial register taking
into account the position of the pixel on the detector and the time needed to cross the
frame store area which is composed of 602 rows. This term is given
by:

\begin{equation}
\Delta t  =  int[(601 + DETY) / 10] \Delta t_{row} + [mod (601 + DETY, 10) +1] \Delta t_{P}
\end{equation}

\noindent
where $int$ indicates the integer part of the number included in the square brackets, $mod (601 + DETY, 10)$ indicates the 
remainder of the division by ten, $\Delta t_{p}$ is the time for a CCD row parallel transfer (15 $\mu$s) 
and $\Delta t_{row}$ (1.7791 ms) is the time interval to read a single row in the serial register:

\begin{equation}
 \Delta t_{row} = (FET - FST)/599
\end{equation}

\noindent
The term $0.5 \Delta t_{row}$ is added to refer the time values calculated by {\sc xrttimetag} to the 
middle of the WT temporal bin (1.7791 ms).

The time-tagging algorithm  requires for each pixel the knowledge of its DETY value, i.e. where 
in the detector Y direction that pixel was illuminated by the X-ray source as it crossed the 
CCD during the columns parallel transfers. 
In other words, it requires the knowledge of the location on the CCD of the X-ray celestial source.
This information is not available in the telemetry data, which stores only 1-D positional information 
along the X direction (DETX). For this reason, the {\sc xrttimetag} module needs 
as inputs the sky coordinates (Right Ascension and Declination) of the pointed X-ray source. 
These sky coordinates are then used, together with the spacecraft attitude file, to 
calculate the corresponding DETY values as a function of time. The time tagging algorithm assumes that 
the field of view is dominated by a bright X-ray source, i.e. that every photon is registered on the 
CCD at the DETY position of this source. This assumption neglects the telescope PSF extension which 
introduces a temporal smearing.

For each XRT observation (respectless of the readout mode) the event arrival times in MET are stored  
in a FITS file where the two keywords MJDREFI and MJDREFF
give the integer and the fractional part of the reference time expressed in TT (Terrestrial Time).
Since the spacecraft clock (set once soon after launch, and left free-running thereafter) 
is expected to drift with time, a correction factor is included in the keyword UTCFINIT
that can be used to correct the time for the clock drift up to the start time of the observation. Furthermore,
a finer correction that accounts also for the drift occurring during the observation can be achieved by using a full list of 
clock offsets that is periodically delivered by the Swift Mission Operation Center and used
as input of the Solar System Barycenter (SSB) 
correction code {\sc barycorr}\footnote{
A full description of  the time assignment in the Swift FITS files is available 
at http://swift.gsfc.nasa.gov/docs/swift/analysis/suppl\_uguide/time\_guide.html.}.  

\section{Observation and data reduction\label{data}}

We have analyzed all the Crab pulsar observations in the Swift archive performed with the XRT
in WT mode and whose observing time interval is covered by a set of  
radio ephemerides provided by
the Jodrell Bank Observatory. 
Observations where the Crab pulsar was more than 5 arcminutes off-axis were excluded
from the analysis because the pulsar is at the edge or out of the WT image strip. 
Table~\ref{log} reports the details on the observations used for the analysis. 
The Level 1 data were downloaded from the HEASARC public data 
archive\footnote{http://heasarc.gsfc.nasa.gov/cgi-bin/W3Browse/swift.pl} and calibrated,
filtered and screened with the {\sc XRTDAS} package included in the HEAsoft 6.11 software release.
Where possible, the observations were also reduced using the attitude file corrected with UVOT 
data\footnote{http://heasarc.gsfc.nasa.gov/lheasoft/ftools/headas/uvotattcorr.html}.
This technique uses UVOT images to provide a more accurate attitude
solution than can be achieved by the Swift star trackers.  Hence it
improves the accuracy of our knowledge of the source position on the XRT
CCD, and hence the accuracy of the assignment of arrival times to the
X-ray photons.

Events with grade 0--2 were selected for the analysis. The source events were extracted from
each observation using a rectangular region with 40 pixel width centered on the pixel with 
higher intensity. This region allows us to select $\sim94\%$ of the PSF of the Crab pulsar, which is located
near the center of the Nebula.
XRT arrival times were referred to the SSB using the Crab
coordinates, RA = 05h 34m 31.972s Dec = 22$^o$ 00' 52''.07 \citep{lyne93}, 
the JPL DE200 solar system ephemeris and the barycentrization code {\sc barycorr}, also transforming the
time reference keywords into TDB.
Due to the Swift low inclination equatorial orbit, and to the optimization strategy adopted for the
pointing planning, each observation comes split into one to several pieces called
snapshots with exposure times 
from a few hundreds of seconds to $\sim 1$ ks, separated by one or more
orbits (each orbit lasts $\sim96$ minutes).

\begin{table}
\scriptsize
\begin{center}
\begin{tabular}{c c c c } \hline
Obs ID         &$P_*~^a$           &$\Delta P_*~^b $         &$P_*-P_{radio}$   \\   
               &    (ms)      & (ms)               & (ns)           \\ \hline  
00050100016    &33.574560903  & 4.5$\times10^{-8}$ &-0.085          \\ 
00058970001    &33.582968895  & 3.4$\times10^{-9}$ & 0.0036         \\ 
00050100019    &33.587278625  & 3.9$\times10^{-7}$ & 0.38           \\ 
00050100020    &33.588669498  & 3.9$\times10^{-8}$ & 0.043          \\ 
00050100021    &33.600921413  & 1.0$\times10^{-7}$ & 0.08           \\ 
00050100025    &33.608537702  & 5.3$\times10^{-8}$ & 0.24           \\ 
80050100026    &33.609195920  & 3.0$\times10^{-8}$ & 0.01           \\ 
00050100027    &33.609251660  & 1.2$\times10^{-8}$ &-0.045          \\ 
00050100029    &33.619760953  & 1.2$\times10^{-7}$ & 0.029          \\ 
00050100030    &33.619760862  & 5.5$\times10^{-8}$ &-0.27           \\ 
00058990008    &33.625457614  & 7.5$\times10^{-8}$ &-0.30           \\ 
00058990009    &33.625663868  & 3.5$\times10^{-7}$ &-1.8            \\ 
00058990010    &33.634050301  & 4.9$\times10^{-8}$ &-0.068          \\ 
00058990011    &33.638754348  & 8.0$\times10^{-8}$ &-0.031          \\ 
00030371012    &33.647613948  & 6.1$\times10^{-8}$ & 0.12           \\ 
00030371013    &33.647655855  & 1.8$\times10^{-8}$ & 0.43           \\ \hline
\end{tabular}
\caption{Results of the pulse phase analysis performed on the observations with at least 3 snapshots.}
\begin{list}{}{}
\item [$^a$]  Best period obtained through the phase analysis technique.
\item [$^b$]  68\% confidence level error on the best period.
\end{list}
\label{phshift}
\end{center}
\end{table}

The Crab pulsar was also observed on 2011 Aug 3 by RXTE (from  55776.95497588 to 55776.96972203 MJD, 
 TDB time scale, 
Obs.ID 96802-01-12-00, with an exposure time of 915 s) following a request of the Swift-XRT hardware team  for 
timing intercalibration. The time range of the RXTE observation was fully covered by 
the Swift-XRT observation (Obs ID 00050100031). We used data  from the Proportional Counter 
Array (PCA, \citealp{bradt93,jahoda06}) collected in Event Mode, time-tagged with a 1 $\mu$s accuracy with respect
to the spacecraft clock, which is maintained to the TT time scale with an absolute time
accuracy of a few $\mu$s.
Data from only four Proportional Counter Units  and from all the layers were analyzed (PCU0 was off). 
Data extraction was performed with the dedicated RXTE tools 
(HEASOFT v.6.11) and standard filtering criteria for obtaining the good-time intervals. We 
filtered the data using the 2--10 keV energy range in order to best match the Swift-XRT energy 
range. Photon arrival times were adjusted to the solar system barycenter using the same
source position as for the Swift data, with the {\sc faxbary} tool, applying the fine clock corrections.

\section{Timing analysis and results\label{analysis} }

\begin{figure}
\centerline{\psfig{figure=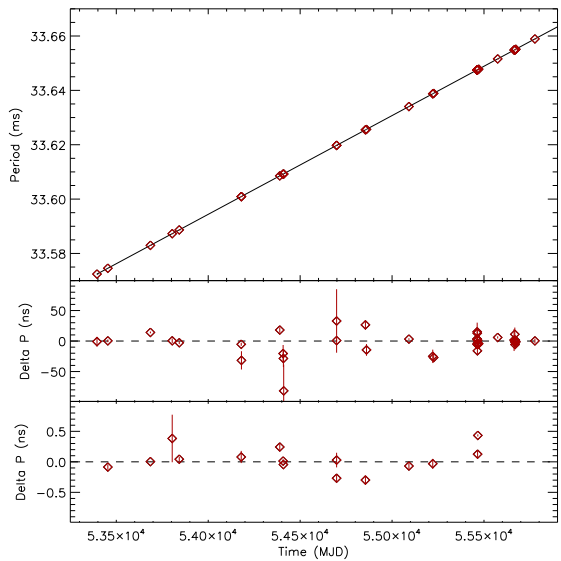,width=8cm,angle=0} }
\caption{ Top panel: best periods obtained for each observation (column 6, Table 1) as weighted mean of the folding analysis 
results performed on single snapshots.
The solid line connects the periods extrapolated at the folding epoch times from the radio ephemeris (column 8, Table 1).
Central panel: residuals between the best periods plotted above and the corresponding periods as
extrapolated from the radio ephemeris. 
Bottom panel: residuals between the periods obtained from the pulse phase analysis (column 2, Table 2) and those extrapolated 
from the radio ephemeris (column 8, Table 1). Due to the Y-axis range in the three panels, errors for several points are too small
to be visible.
\label{period}}
\end{figure}

\subsection{Accuracy of the Crab period estimation}

For each observation we have derived the best Crab period with the following procedure:
\begin{enumerate}
\item We have applied a folding analysis (e.g. \citealp{lorimer04}) to the barycentered arrival times in each snapshot,
fixing the  folding epoch time to the observation central time (column 5 in Table~\ref{log}), and 
searching in a period range centered on the expected period (column 8 in Table~\ref{log}) 
evaluated using the Crab radio ephemeris.
The period search was performed with a step resolution $\Delta P=P^2/(N\Delta T)$, where {\it N}=100 is the number 
of phase bins used to sample the pulse profile and $\Delta T$ is the data time span for each snapshot.
\item For each snapshot periodogram we derived the best period by fitting the 
$\chi^2$ peak with a Gaussian function. The error on the best period (68\% confidence level) was 
evaluated computing the period 
interval corresponding to a unit decrement with respect to the maximum in the $\chi^2$ curve \citep{cusumano03} 
i.e. $\Delta P =P(\chi^2_{max}) - P(\chi^2_{max} - 1)$ .
\item The best period for the whole observation (columns 6, 7 in Table~\ref{log}) was obtained as the 
weighted mean of the best period values of the snapshots, using  the errors on each period as weight. 
\end{enumerate}

Figure~\ref{period} (top panel) shows the best period values obtained for each observation vs. time 
and compares them with the ones derived from the radio ephemeris.
The central panel shows the residuals with respect to the values extrapolated from the radio ephemeris (see also
column 10 in Table~\ref{log}). 
The folding analysis on these observations allows us to obtain an accuracy of the order of a few to tens of ns in 
the determination of the period, depending on the data time span and on the frequency of the snapshots within
each observation.

The precision of the timing parameters can be enhanced by performing a pulse phase analysis  on
the observations with several snapshots (at least 3). This consists in maintaining a phase pulse coherence over the entire time 
interval covered by the observation. We have folded the event arrival times of each snapshot with 
the same epoch time used in the previous analysis and with the observation best period 
(column 6 in Table~\ref{log}). Phase shifts of the pulse profile are expected because of the
indetermination of the period within its error and of the presence of period derivatives. 
These phase shifts were evaluated by measuring the position of the main peak fitting it with
a Lorentzian
function plus a constant. The width of the Lorentzian was fixed to the value obtained in the 
observation with the highest statistics (ObsID 00058970001). As the main peak has an 
asymmetric shape, we checked for the presence of systematics in the determination of the peak
position due to the choice of the model: fits with a Gaussian function and with a
parabola agree within errors with the values obtained with the Lorentzian model, that were 
therefore used throughout the analysis.

In the absence of frequency irregularities (glitches), the correction to the frequency and its derivative 
can be obtained by fitting the phase lags by a third degree model:
\begin{equation}
\Delta\phi(t)=\Delta\nu_0(t-t_0)+\frac{1}{2}\Delta\dot\nu_0(t-t_0)^2+\frac{1}{6}\Delta\ddot\nu_0(t-t_0)^3
\end{equation}
where $\Delta\phi$ is the measured phase difference at time $t$, $t_0$ is the epoch time, and $\Delta\nu_0$,
$\Delta\dot\nu_0$, and $\Delta\ddot\nu_0$ are the correction to the frequency and to its first and second derivative,
respectively. 
The data time span of the Crab observations is not long enough for the second derivative of the frequency to 
contribute significantly to the phase shift. Therefore the third degree term of the above polynomial is ignored in the fit.
Table~\ref{phshift} reports the periods P$_*$ (column 2) obtained applying $\Delta\nu_0$
from  Eq. 4 to the periods 
reported in Table 1, and their associated errors $\Delta P_*$ (column 3) derived as $\sigma_{\Delta\nu_0}\times
P_*^2$, where $\sigma_{\Delta\nu_0}$ is the 68\% confidence level of $\Delta\nu_0$. Differences with respect to the 
periods extrapolated from the radio ephemeris are reported in column 4. 
The latter are also plotted in the bottom panel in Figure~\ref{period}.
The pulse phase analysis allows us to obtain an accuracy of the order of tens of picoseconds in 
the determination of the period, depending on the data time span, down to $\sim 4$ ps 
($\Delta P/P \sim 10^{-9}$) for the observation with 
the longest time span (ObsID 00058970001).
The accuracy on the estimate of the first derivative of the period depends both on the length of the data time span and on
the number of snapshots; for ObsID 00058970001 we find a deviation of the period first derivative with respect to 
the value extrapolated from the radio ephemeris of 0.1\%.
\vspace{0.6cm}
\begin{figure}
\centerline{\psfig{figure=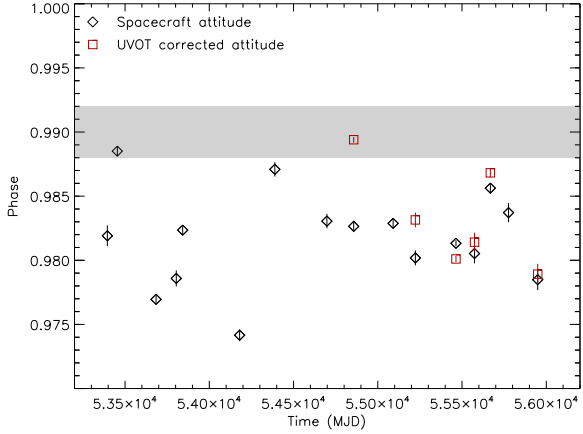,width=8cm,angle=0} }
\caption{ Phase position of the Crab pulsar main peak (black diamonds). 
The shaded strip marks the phase range where the main peak is 
measured by RXTE-PCA. Red squares correspond to the phase position measured after correcting the attitude file
with UVOT data. Observations with the same radio
ephemeris sample interval were folded together and are thus represented by a single point. 
\label{phase}}
\end{figure}

\begin{figure}
\centerline{\psfig{figure=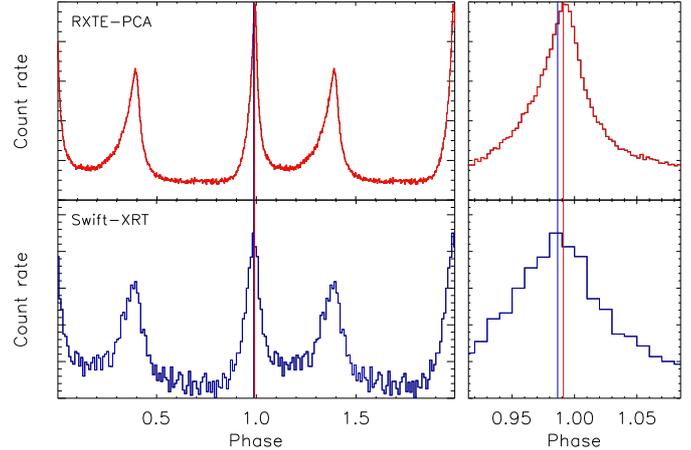,width=9cm,angle=90} }
\caption{Comparison between the Crab pulse profile obtained with RXTE-PCA (top panels) and that obtained
with the Swift-XRT ObsID 00050100031. The count rates in the Y-axis are in arbitrary
units. Vertical lines mark the position of the main peak (red: RXTE-PCA; blue:
Swift-XRT). The right panels show a zoom of the main peak interval.
\label{peak}}
\end{figure}

\begin{table}
\scriptsize
\begin{tabular}{l l l l l }
Instrument & Obs ID   & $T_{start}$ (TDB)    & $T_{elapsed}$  &Exposure (s)\\   
           &          &  (MJD)               &  (s)           &  (s) \\  \hline 
XRT-PD &    00050100010  & 53396.14367194747 & 59153.3   &15326.4  \\
       &    00050100014  & 53399.01104749178 & 134287.2  &29743.9  \\
       &    00050101002  & 53414.08223807389 & 750.5	 &750.5\\ 
       &    00050102002  & 53414.15029296864 & 692.2	 &692.2  \\ 
       &    00050101003  & 53426.60594507941 & 2333.8	 &2333.8 \\ 
       &    00050101004  & 53437.05995839463 & 40799.0   &3697.9 \\ 
       &    00050100015  & 53442.21581378842 & 2354.5	 &2354.5 \\ 
\vspace{0.3 cm}
       &    00050100018  & 53470.36073094432 & 806.5	 &806.5  \\ 
BAT    & 00050101001	& 53414.08275463	&  250.198	&   250.1  \\
       & 00050101002	& 53414.08831019	&  250.189	&   250.1  \\
       & 00050101003	& 53426.60660880	&  1250.255	&   1690.0  \\
       & 00050102003	& 53438.00515046	&  100.20	&   23146.0  \\
       & 00050100016	& 53454.07055556	&  1500.32	&   1500.3  \\
       &  00050100021	&  54179.53190972	&   1727.93	&    13173.1  \\
       &  00030371015	&  55471.00218500695	&   76.599	&    76.5  \\ 
       &  00306746000	&  54544.09232639	&   696.857	&    1142.0  \\
       &  00050100029	&  54698.00137482176	&   283.446	&    34621.8   \\
       &  00058990008	&  54855.21901834028	&   293.189	&    11641.0  \\
       &  00058990009	&  54860.77601602546	&   755.000	&    755.0  \\
       &  00058990012	& 55574.59433546990	&  924.700	&   924.7  \\
       &  00050100031	& 55776.96256000000	&  970.165	&   970.1  \\ \hline
\end{tabular}
\caption{XRT-PD and BAT Crab observations. \label{lr-bat}}

\end{table}

\begin{figure}
\centerline{\psfig{figure=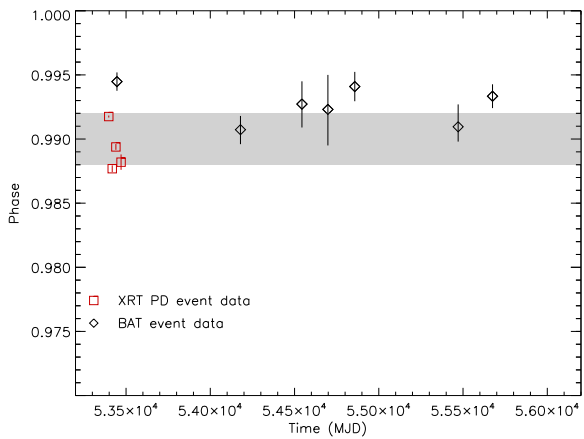,width=8cm,angle=0} }
\caption{
Phase position of the Crab pulsar main peak for XRT-PD (red squares) and BAT (black diamonds) observations.
The shaded strip marks the phase range where the main peak is measured by RXTE-PCA. For each instrument, 
observations with the same radio
ephemeris sample interval were folded together and are thus represented by a single point. 
\label{peak_lr_bat}}
\end{figure}

\vspace{0.6cm}

\subsection{Absolute timing accuracy}

The absolute timing of XRT in WT mode was evaluated measuring the phase position of the main pulse peak
in the Crab profile obtained folding the XRT data with the radio ephemeris. 
Observations performed within the same radio ephemeris sample interval were folded together to improve the 
statistics of the pulse profile. 
As for the phase analysis, the phase of the main peak was then determined for each profile by fitting it with 
a Lorentzian function plus a constant.
Figure~\ref{phase} (black diamonds) shows the phase of the main peak vs. time, together with the range of
values measured by RXTE \citep{rots04}. The positions of the main peak for each XRT
observation are not consistent with
each other and range between 0.9742 and  0.9885. The pulse profile leads the radio profile and also arrives
systematically earlier with respect to the RXTE measure. 
The average position of the main peak in
XRT is at radio phase $0.9817\pm0.0005$, corresponding to $\sim 620 \mu$s earlier than the radio peak 
and $\sim 270 \mu$s earlier than RXTE.  

This analysis was also repeated for the observations reduced with the attitude file
corrected with UVOT data (see Section 3 and first column in Table 1). 
Figure~\ref{phase} (red squares) shows the result: the average
values of the main peak phase is at $0.9854\pm0.0002$, corresponding to  $\sim490 \mu$s earlier than the 
radio peak and $\sim 150 \mu$s earlier than RXTE, with a scatter ranging between 0.9801 and 0.9894.

The absolute phase was also verified using the RXTE observation performed simultaneously with Swift-XRT
ObsID 00050100031 (for which only the spacecraft attitude file was available). 
Given their better time resolution, the RXTE data were folded using 400 phase bins.
The XRT data were restricted to the energy range between 2 and 10 keV to better match the RXTE 
energy selection.
The two pulse profiles, obtained folding the data with the radio ephemeris, are shown in 
Figure~\ref{peak}. The comparison between the two profiles shows that the position of the main peak 
observed in XRT is  earlier than the RXTE profile with an advance of only $200\pm38$ $\mu$s.
We have verified that the phase difference is not due to the different spectral content of the two datasets because of the
different shape of the effective area in the two instruments: using the observation with the longest time data span 
(ObsID 00058970001) we do not observe any significant variation of the main peak phase when selecting in different energy
range (2--10 keV, 4--10 keV, 6--10 keV, 8--10 keV). We have also verified that the phase difference is not due to the
worse time resolution of the XRT dataset with respect to the RXTE one: we have modified the arrival times  of the
RXTE events by resampling them with a time resolution of 1.7791 ms and corrected them to the solar system barycenter;
the resulting folded profile shows both a broadening ($\sim 35\%$) and a shift ($\sim 35 \mu$s in advance) 
of the main peak. However, this shift is too small to account for the phase
difference observed in the XRT profile.

We have checked if the difference in absolute timing observed between RXTE and XRT data in WT mode could also be observed in XRT data
collected in PD mode (with 140 $\mu$s time resolution) and BAT data collected in event mode (with 100 $\mu$s time resolution). We have
analyzed all Crab observations with suitable statistics available in the Swift archive. Table~\ref{lr-bat} 
reports the details on the observations used for this
analysis. We selected PD data in the 2--10 keV range and BAT data in the 15--150 keV range. The barycentered data were 
analyzed using the same procedure as for the WT data, and the main peak positions obtained with a Lorentzian fit are plotted in 
Figure~\ref{peak_lr_bat}. The main peak position obtained by the pulse profiles of the PD mode data are fully consistent with the RXTE
result; also for the BAT data we find a good agreement with RXTE. Moreover, for both data sets, the scattering is significantly smaller than
that observed in XRT in WT mode.

\section{Conclusions\label{conclusion}}
We have analyzed  the set of Swift-XRT Crab observations in order to qualify the timing performance of the XRT
observing in WT mode. \\
The accuracy in the Crab period determination is mainly limited by the time span covered by the observation
with a minimum difference of a few picoseconds with respect to the period extrapolated from radio ephemeris
for the longest XRT observation.\\
We observe a displacement in phase of 
the Crab pulse profile with respect to the profile observed 
by RXTE-PCA: 
the main peak observed in XRT leads the RXTE-PCA one by $\sim 270 \mu$s
using the spacecraft attitude files, or $\sim 150 \mu$ s when UVOT 
corrected attitude files are available.
This represents a 15\% (9\% using UVOT attitude files) shift in absolute timing compared to the
time resolution of the XRT in WT mode. We have verified that this shift is not due to a spectral
effect, nor to the worse time resolution of the XRT dataset with respect to the RXTE one. 
We observe a significant scatter in the position of the main peak among the XRT
observations. 
Conversely, the peak positions obtained by the analysis of the Crab observations collected by XRT in PD mode and 
by BAT in event mode are in good agreement with the RXTE measurements. This suggests that the discrepancy observed in XRT WT mode
is not due to the Swift spacecraft clock, but most likely to a systematic offset in the time
assignment specific to this mode.  However, the precision of the pulse period determination in WT mode suggests 
that the phase offset is stable within each observation even if it is not between observations. 

\begin{acknowledgements}
The authors wish to thank the anonymous referee for critical review which allowed us
to significantly improve the paper.
APB and JPO acknowledge the support of the UK Space Agency.
This work has been supported by ASI grant I/011/07/0.
\end{acknowledgements}

\bibliographystyle{aa}

\begin{thebibliography}{}

\bibitem[Abbey et al.(2006)]{abbey06} Abbey, T., Carpenter, J., 
Read, A., et al.\ 2006, The X-ray Universe 2005, 604, 943 
\bibitem[Barthelmy et al.(2005)]{barthelmy05} Barthelmy, S.~D., 
Barbier, L.~M., Cummings, J.~R., et al.\ 2005, \ssr, 120, 143 
\bibitem[Bradt et al.(1993)]{bradt93} Bradt, H.~V., Rothschild, R.~E., 
\& Swank, J.~H.\ 1993, \aaps, 97, 355 
\bibitem[Burrows et al.(2005)]{burrows05} Burrows, D.~N., Hill, 
J.~E., Nousek, J.~A., et al.\ 2005, \ssr, 120, 165 
\bibitem[Cusumano et al.(2003)]{cusumano03} Cusumano, G., Massaro, E., \& Mineo, 
T.\ 2003, \aap, 402, 647 
\bibitem[Gehrels et al.(2004)]{gehrels04} Gehrels, N., 
Chincarini, G., Giommi, P., et al.\ 2004, \apj, 611, 1005
\bibitem[Hill et al.(2004)]{hill04} Hill, J.~E., Burrows, 
D.~N., Nousek, J.~A., et al.\ 2004, \procspie, 5165, 217 
\bibitem[Jahoda et al.(2006)]{jahoda06} Jahoda, K., Markwardt, 
C.~B., Radeva, Y., et al.\ 2006, \apjs, 163, 401 
bibitem[Kirsch et al.(2004)]{kirsch04} Kirsch, M.~G.~F., 
Becker, W., Benlloch-Garcia, S., et al.\ 2004, \procspie, 5165, 85 
\bibitem[Kirsch et al.(2004)]{kirsch04} Kirsch, M.~G.~F., 
Becker, W., Benlloch-Garcia, S., et al.\ 2004, \procspie, 5165, 85 
\bibitem[Kuster et al.(2002)]{kuster02} Kuster, M., Kendziorra, 
E., Benlloch, S., et al.\ 2002, arXiv:astro-ph/0203207 
\bibitem[Lorimer et al.(2004)]{lorimer04} Lorimer, D.~R., Kramer, 
M., Ellis, R., et al.\ 2004, Handbook of pulsar astronomy, by D.R.~Lorimer 
and M.~Kramer.~Cambridge observing handbooks for research astronomers, 
Vol.~4.~Cambridge, UK: Cambridge University Press, 2004,  
\bibitem[Lyne et al.(1993)]{lyne93} Lyne, A.~G., Pritchard, 
R.~S., \& Graham-Smith, F.\ 1993, \mnras, 265, 1003 
\bibitem[Molkov et al.(2010)]{molkov10} Molkov, S., Jourdain, 
E., \& Roques, J.~P.\ 2010, \apj, 708, 403 
\bibitem[Roming et al.(2005)]{roming05} Roming, P.~W.~A., 
Kennedy, T.~E., Mason, K.~O., et al.\ 2005, \ssr, 120, 95 
\bibitem[Rots et al.(2004)]{rots04} Rots, A.~H., Jahoda, K., 
\& Lyne, A.~G.\ 2004, \apjl, 605, L129 
\bibitem[Terada et al.(2008)]{terada08} Terada, Y., Enoto, T., 
Miyawaki, R., et al.\ 2008, \pasj, 60, 25 

\end{thebibliography}
{}

\end{document}